\documentclass[twocolumn,showkeys,aps,prb,showpacs]{revtex4-1}
\usepackage{graphicx}
\usepackage[CJKbookmarks,dvipdfm,colorlinks,linkcolor=blue,citecolor=blue]{hyperref}

\begin{document}

\title{Phonon transport of three-fold degeneracy topological semimetal MoP}

\author{San-Dong Guo}
\affiliation{School of Physics, China University of Mining and
Technology, Xuzhou 221116, Jiangsu, China}

\begin{abstract}
Recently, three-component new fermions in topological semimetal MoP are  experimentally observed, which may have potential applications like topological qubits, low-power electronics and spintronics.  These are closely related to  thermal transport properties of MoP.
 In this work, the phonon transport  of  MoP is investigated  by solving  the linearized phonon Boltzmann equation within
the single-mode relaxation time approximation (RTA).
The calculated room-temperature lattice thermal conductivity  is  18.41  $\mathrm{W m^{-1} K^{-1}}$  and 34.71 $\mathrm{W m^{-1} K^{-1}}$ along the in-  and cross-plane directions, exhibiting very strong anisotropy.  The isotope and size effects on the lattice thermal conductivity are also considered. It is
found that isotope scattering produces little effect, and phonon has little contribution  to the lattice thermal conductivity, when  phonon mean free path(MFP) is larger than 0.15 $\mathrm{\mu m}$ at 300 K.
It is noted that average room-temperature lattice thermal conductivity of MoP is  lower than that of representative Weyl semimetal
TaAs, which is due to smaller group  velocities and larger Gr$\mathrm{\ddot{u}}$neisen parameters.  Our works provide  valuable informations for the thermal management of MoP-based nano-electronics devices, and motivate  further experimental works to study thermal transport of MoP.

\end{abstract}

\pacs{72.15.Jf, 71.20.-b, 71.70.Ej, 79.10.-n ~~~~~~~~~~~~~~~~~~~~~~~~~~~~~~~~~~~Email:guosd@cumt.edu.cn}

\maketitle

\section{Introduction}
The study of  new type of topological nontrivial phase, from topological insulator to semimetal,
has attracted great research interest in recent years\cite{q1,q2,q3,q4,q5,q5-1,q6,q7,q8}.
Dirac semimetals, Weyl semimetals and nodal line semimetals are three kinds of  representative topological semimetals\cite{q4,q5,q8}.  By including
the spin-orbit coupling (SOC) or mass term, the nodal line structure can
turn into Weyl points,  Dirac points  or  a topological insulator\cite{q9}.  The $\mathrm{Na_3Bi}$ is a classic  Dirac Semimetal, which has been confirmed by angle-resolved photoemission spectroscopy (ARPES), and three-dimensional (3D) Dirac fermions with linear dispersions have been detected  along all momentum directions\cite{q5-1}. The first experimental realization
of a Weyl semimetal is in TaAs, and both the Weyl fermions
and Fermi arc surface states have been detected\cite{q4}, which has been further confirmed by  ARPES\cite{q10}.
Both Weyl  and Dirac fermions are even-component, which are  two- and four-component, respectively. Recently, three-component fermions have been  predicted  in ZrTe with the WC-structure   by first-principles calculations\cite{q11}.  Then, a triply degenerate point in  MoP with the WC-structure  has been detected by ARPES, and pairs of Weyl points  coexist with the three-component fermions\cite{q7}.

Most studies  focus on the topological electronic
structure of these exotic materials, but little research can be found  to investigate the thermal transport in
theses topological semimetals. As is
well known,   the thermal transport property is closely related to the application of a material in nano-devices. Recently,
thermal transport in TaAs has been investigated from a first principles calculation, and the lattice thermal conductivity  shows  obvious anisotropy
along the a(b) and c crystal axis\cite{q12}. The thermoelectric properties of  TaAs have also been investigated, and the maximum thermoelectric figure of merit $ZT$ reaches  up to 0.63 at 900 K in n-doping along  c crystal axis\cite{q13}. Here,  the phonon transport properties of three-fold degeneracy topological semimetal MoP are investigated by solving the phonon Boltzmann transport equation
. The lattice thermal conductivity is predicted, and the isotope and size effects on the lattice thermal conductivity are also studied.
The lattice thermal conductivity shows a distinct anisotropic  property along the a(b) and c crystal axis.
The phonon mode group  velocities, lifetimes and  Gr$\mathrm{\ddot{u}}$neisen parameters
are calculated to understand deeply  the phonon transport  of MoP.

\begin{figure}
  \includegraphics[width=7cm]{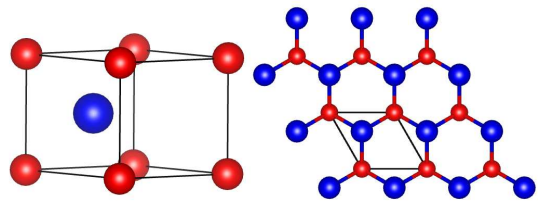}
  \caption{(Color online)The crystal structure of MoP in one unit cell, and  top view of the crystal structure. The blue and red balls represent Mo and P atoms, respectively.}\label{st}
\end{figure}

The rest of the paper is organized as follows. In the next
section, we shall give our computational details. In the third section, we shall present phonon transport of MoP. Finally, we shall give our conclusions in the fourth section.

\section{Computational detail}
First-principles calculations are performed within the projected augmented wave (PAW) method, and the exchange-correlation energy functional of generalized gradient approximation of the Perdew-Burke-Ernzerhof (GGA-PBE) is adopted,
as implemented in the VASP code\cite{pv1,pv2,pbe,pv3}.
A plane-wave basis set is employed with
kinetic energy cutoff of 400 eV, and the $s^2p^3$ orbitals of P  and  $4p5s4d$ orbitals of Mo  are treated as valance ones.
The electronic stopping criterion is $10^{-8}$ eV.
The  lattice thermal conductivity of  MoP  is carried out by solving linearized phonon Boltzmann equation with the single mode RTA,   as implemented in the Phono3py code\cite{pv4}. The interatomic force constants (IFCs) are calculated by
the finite displacement method.
 The second-order harmonic IFCs
are calculated using a 4 $\times$ 4 $\times$ 4  supercell  containing
128 atoms with k-point meshes of 2 $\times$ 2 $\times$ 2. Using the harmonic IFCs, phonon dispersion of MoP can be calculated by  Phonopy package\cite{pv5}.  The phonon dispersion determines the allowed three-phonon scattering processes, and further the  group velocity  and specific heat can be attained. The third-order anharmonic IFCs are calculated using a 3 $\times$ 3 $\times$ 3
supercells containing 54 atoms with k-point meshes of 3 $\times$ 3 $\times$ 3.
 Based on third-order anharmonic IFCs, the three-phonon scattering rate can be calculated, and  further   the phonon lifetimes can be attained. To compute lattice thermal conductivities, the
reciprocal spaces of the primitive cells  are sampled using the 20 $\times$ 20 $\times$ 20 meshes.
\begin{figure}
  \includegraphics[width=8cm]{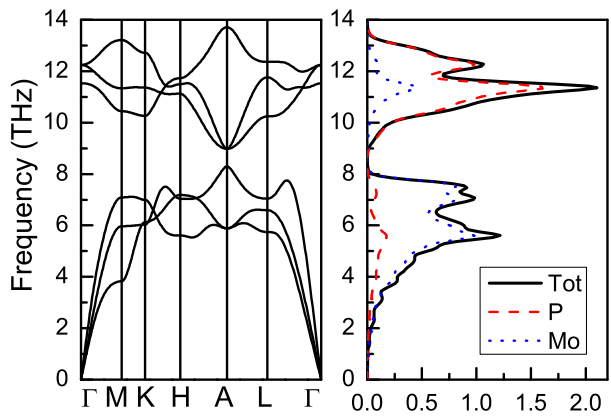}
  \caption{(Color online)Phonon band structures of MoP with the
corresponding total density of states (DOS) and  atom partial DOS (PDOS). }\label{ph}
\end{figure}

\begin{figure}
  \includegraphics[width=8cm]{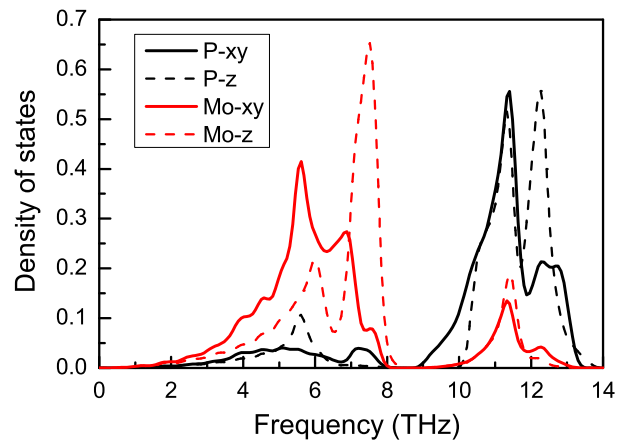}
  \caption{(Color online)The x(y) and z components of atom PDOS of MoP.}\label{ph1}
\end{figure}
\begin{figure*}[!htb]
  \includegraphics[width=15cm]{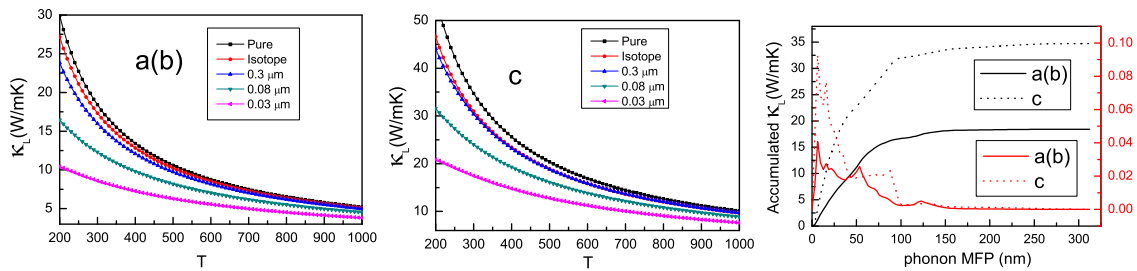}
  \caption{(Color online) The lattice thermal conductivities  of infinite (Pure and Isotope) and finite-size (0.3, 0.08 and 0.03 $\mathrm{\mu m}$) MoP as a function of temperature, including a(b)  and c directions; The cumulative lattice thermal conductivity (300 K) of  infinite (Pure) MoP with respect to phonon mean free path, and the derivatives.}\label{kl}
\end{figure*}

\begin{figure}
  \includegraphics[width=7.0cm]{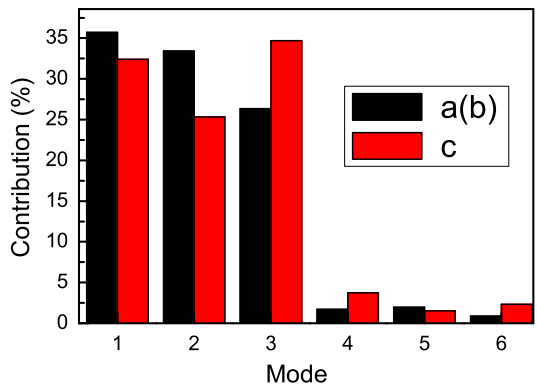}
  \caption{(Color online) The phonon modes contributions along a(b) and c directions to total lattice thermal conductivity at room temperature. 1, 2, 3 represent TA1, TA2 and LA branches and 4, 5, 6 for optical branches.}\label{mkl}
\end{figure}

\section{MAIN CALCULATED RESULTS AND ANALYSIS}
 As shown in \autoref{st},  MoP  possesses  WC-type crystal structure with space group  $P\bar{6}m2$ (No. 187), where Mo and
P atoms occupy the 1d (1/3, 2/3, 1/2) and 1a (0,0,0) Wyckoff positions, respectively. The experimental lattice constants ($a$=$b$=3.231 $\mathrm{{\AA}}$, $c$=3.207 $\mathrm{{\AA}}$ ) are used to investigate elastic properties and phonon transport of MoP\cite{q14}.
The five independent elastic  constants $C_{ij}$  are calculated, and the corresponding values (in
GPa): $C_{11}$=359.00, $C_{12}$=153.73, $C_{13}$=160.14,
$C_{33}$=515.15  and  $C_{44}$=169.22.
The hexagonal mechanical stability criteria are given by\cite{q15}: $C_{44}$$>$0, $C_{11}$$>$$|C_{12}|$ and ($C_{11}$+2$C_{12}$)$C_{33}$$>$2$C_{13}^2$.
By simple calculation, these criteria are satisfied for MoP, proving mechanical stability of MoP. Based on  elastic  constants $C_{ij}$, the bulk ($B$), shear ($G$), and Young ($E_{xx}$ and $E_{zz}$) moduli  (in
GPa) are 239.10, 134.96, 274.28  and 415.11, respectively.

Based on the eigenvalues of harmonic IFCs matrix, the phonon dispersion of
 MoP along several high symmetry paths is obtained.  The phonon dispersion along with total and atom partial density of states (DOS) are plotted in \autoref{ph}. It is clearly seen that no imaginary frequencies are produced in the phonon dispersion of
 MoP, which indicates the thermodynamic stability of MoP.
There are 3 acoustic and 3 optical phonon branches because of two atoms  per unit cell. It can be noted  that
there is  a phonon band gap of 0.68 THz, separating acoustic branches from optical branches.
 The phonon band gap may be due to different atomic masses of P and Mo atoms\cite{m1,m3}.
 From atom partial DOS, acoustic (optical) branches
are mainly contributed by the vibrations of
Mo (P) atoms. According to \autoref{ph1}, the x(y) modes of P and Mo atoms  have stronger coupling with the z modes in high frequency region than in low frequency region. As is well known, the phonons in the low frequency region dominate the lattice thermal conductivity. Therefore, the weak  coupling between x(y) and z modes may lead to strong anisotropy of lattice thermal conductivity along a(b) and c directions.

Based on harmonic and anharmonic IFCs, the intrinsic lattice thermal conductivity of MoP  is calculated by solving the linearized phonon Boltzmann equation within single-mode RTA method.
The lattice thermal conductivities  of infinite (Pure)  MoP along a(b) and c directions as a function of temperature are plotted in \autoref{kl}.
In the considered temperature region, the
lattice thermal conductivity of MoP  decreases with increasing temperature,  which is due to
intrinsic enhancement of phonon-phonon scattering.
It is found that the lattice thermal conductivity
 exhibits obvious anisotropy, in which the lattice thermal conductivity
along  c direction  was higher than that along  a(b)direction.
The room-temperature lattice thermal conductivities of infinite (Pure) MoP along a(b) and c directions are 18.41  $\mathrm{W m^{-1} K^{-1}}$ and  34.71  $\mathrm{W m^{-1} K^{-1}}$, respectively. An anisotropy factor\cite{q12},  defined as $\eta=(\kappa_{L}(cc)-\kappa_{L}(aa))/\kappa_{L}(aa)$, is used to measure the anisotropic strength, and the corresponding value is 88.5\%.  The lattice thermal conductivity is related to Young's modulus by $\kappa_L\sim \sqrt{E}$\cite{q16}.  It is found that the Young's modulus along a(b) direction is smaller than that along c direction, which is identical with that of lattice thermal conductivity.

Next, we consider phonon-isotope scattering and boundary scattering. Based on the formula proposed  by Shin-ichiro Tamura\cite{q24}, the phonon-isotope scattering  is  calculated. For boundary scattering,   the scattering rate can be simply calculated by $v_g/L$, in which $v_g$, $L$  are the group velocity and  the boundary mean free path, respectively.  The lattice thermal conductivities  of infinite (Isotope) and finite-size (0.3, 0.08 and 0.03 $\mathrm{\mu m}$) MoP as a function of temperature are shown in \autoref{kl}. At 300 K, the lattice thermal conductivities along a(b) and c directions are 17.34 $\mathrm{W m^{-1} K^{-1}}$  and 31.01 $\mathrm{W m^{-1} K^{-1}}$ for infinite (Isotope)  MoP, which are slightly lower than those of infinite (Pure)  MoP.
With the sample size decreasing, it is clearly seen that the lattice thermal conductivity decreases because of enhanced boundary scattering.
For the 0.3, 0.08 and 0.3 $\mathrm{\mu m}$ cases, at room temperature, the lattice thermal conductivity  is reduced by about 13.25\%, 33.27\% and 53.31\%  with respect to that of infinite (pure) case along a(b) direction, and 12.47\%, 30.79\% and 49.63\%  along c direction.
The contribution to  total lattice thermal
conductivity from individual phonon modes with different MFP  can predict the
size effect, because it shows how phonons with
different MFP contribute to the thermal conductivity.
The cumulative lattice thermal conductivity along with the derivatives with respect to MFP (300 K)  along a(b) and c directions are shown in \autoref{kl}.
 It is clearly seen that the cumulative lattice thermal conductivity along both a(b) and c directions approaches saturation value  with MFP increasing.
 Phonons with MFP larger than 0.15 $\mathrm{\mu m}$ only lightly contribute  to the lattice thermal conductivity.
 Phonons with MFP smaller than 0.04 $\mathrm{\mu m}$ along a(b) direction and 0.03 $\mathrm{\mu m}$ along z direction
 contribute  around half  to the lattice thermal conductivity. According to the derivatives, phonons  dominating  the lattice thermal conductivity   along c direction have  shorter MFP than ones along a(b) direction.
\begin{figure*}
  \includegraphics[width=15.2cm]{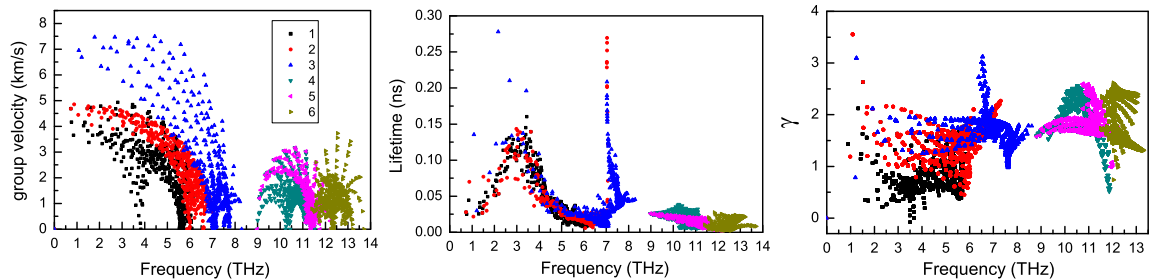}
  \caption{The mode level phonon group velocities, phonon lifetimes (300K) and   Gr$\mathrm{\ddot{u}}$neisen parameters  of infinite (Pure) MoP in the first Brillouin zone. 1, 2, 3 represent  TA1, TA2 and LA branches and 4, 5, 6 for optical branches.}\label{v}
\end{figure*}

At room temperature, the phonon modes contributions  along a(b) and c directions to the total lattice
thermal conductivity are plotted in \autoref{mkl}. It is evident that the  acoustic phonon modes
 dominate  the lattice thermal conductivity of MoP, while the contribution from the optical branches  is quite small.
The acoustic branches provide a contribution of 95.48\% along a(b) direction and 92.41\% along c direction.
Along a(b) direction, any of  two transverse acoustic phonon modes (TA1 or TA2) has  larger contribution than longitudinal acoustic phonon mode
(LA), while LA provides larger contribution than TA1 or TA2 along c direction.
To gain more insight into phonon transport  of MoP,  the mode level phonon group velocities
and lifetimes are plotted in \autoref{v}.
The largest  group velocities   of  TA1 and TA2 branches near $\Gamma$ point are almost the same, and the corresponding value is  4.68  $\mathrm{km s^{-1}}$.
It is 6.95 $\mathrm{km s^{-1}}$ for the largest  group velocity    of  LA branch near $\Gamma$ point.  The most of group
velocities of LA branch are larger than those of TA2 and TA1 branches, while the
smallest group  velocities are observed for the TA1 branch. It  can be noted that  the most of group
velocities of  acoustic branches are higher than those of optical branches.
It is found that most of   phonon lifetimes of three acoustic  branches  overlap below about 6.5 THz.
Unexpectedly,  the phonon lifetimes of  TA2 and LA  near 7.5 THz become large.
It is also noted that  most of  phonon lifetimes
of  three acoustic phonon modes are larger than those of
optical branches. Together with their large group velocities,
 these results lead to the dominant
contribution from the three
acoustic phonon modes  to the total lattice thermal conductivity.
Mode Gr$\mathrm{\ddot{u}}$neisen parameters can provide information about anharmonic interactions, determining the intrinsic phonon-phonon scattering,
and can be attained directly from the third
order anharmonic IFCs.  The mode level  Gr$\mathrm{\ddot{u}}$neisen parameters of  infinite (Pure) MoP are shown in \autoref{v}.
The large  $\gamma$ is in favour of low  lattice thermal conductivity because of strong anharmonicity.
For TA2, LA and optical  branches, the
$\gamma$ is fully positive. However, TA1 branch has  a small negative $\gamma$. The average  Gr$\mathrm{\ddot{u}}$neisen parameter is 1.57, indicating relatively strong anharmonic phonon scattering.

\section{Conclusion}
Phonon transport  in Weyl semimetal TaAs has been investigated by combining first principle calculations
and Boltzmann phonon transport equation\cite{q12,q13}.  The lattice thermal conductivity of TaAs also shows  obvious anisotropy along a(b) and c directions.
The room-temperature one is 39.26  $\mathrm{W m^{-1} K^{-1}}$ along a(b) direction, which is larger than 24.78 $\mathrm{W m^{-1} K^{-1}}$ along c direction\cite{q12}. This is different from that of MoP, where the lattice thermal conductivity along a(b) direction is smaller than one along c direction.
It is found that the average  lattice thermal conductivity ($\kappa_L(av)$=($\kappa_L(aa)$+$\kappa_L(bb)$+$\kappa_L(cc)$)/3) of MoP is smaller than that of TaAs.  This can be explained by group velocities and  Gr$\mathrm{\ddot{u}}$neisen parameters.   Calculated results show MoP has smaller group velocities than TaAs\cite{q13},  which partially gives rise to lower lattice thermal conductivity for MoP than TaAs. Another factor is that MoP has larger Gr$\mathrm{\ddot{u}}$neisen parameters than TaAs\cite{q13}, which induces more stronger anharmonicity, leading to lower  lattice thermal conductivity.

In summary, based on the first-principles calculations and semiclassical Boltzmann transport theory, we investigate the phonon transport properties of MoP,  which contains three-component fermions in the bulk electronic structure.
The  lattice thermal conductivity of MoP shows an obvious anisotropy, and
the room-temperature lattice thermal conductivity  is  18.41  $\mathrm{W m^{-1} K^{-1}}$  and 34.71 $\mathrm{W m^{-1} K^{-1}}$
 along the a(b) and c crystal axis, respectively. For designing nanostructures,
 the size dependence of lattice thermal conductivity  is investigated, and  phonons with MFP larger than 0.15 $\mathrm{\mu m}$  have little contribution to the total lattice thermal conductivity.
The lower  lattice thermal conductivity of MoP than TaAs
 can be understood by lower group velocities and larger Gr$\mathrm{\ddot{u}}$neisen parameters.   Our works shed light on phonon transport of MoP, and will motivate farther experimental studies of thermal transport properties of  topological semimetals.

\begin{acknowledgments}
This work is supported by the National Natural Science Foundation of China (Grant No.11404391). We are grateful to the Advanced Analysis and Computation Center of CUMT for the award of CPU hours to accomplish this work.
\end{acknowledgments}

\end{document}